\documentstyle[prd,epsfig,preprint,aps,tighten]{revtex}
\def \gsim {\mbox{${}^> \hspace*{-7pt} _\sim$}}

\begin{document}
\tightenlines
\begin{titlepage}
\preprint{
\vbox{\hbox{IFIC/00-72}
      \hbox{hep-ph/0011245}}}
\title{Four--neutrino Oscillations at SNO}
\author{
  M.~C.~Gonzalez-Garcia\thanks{E-mail: \tt concha@flamenco.ific.uv.es} and
  C.~Pe\~na-Garay\thanks{E-mail: \tt penya@flamenco.ific.uv.es},
  }
\address{
  Instituto de F\'{\i}sica Corpuscular~--~C.S.I.C., Universitat de Val\`encia, \\
  Edificio Institutos de Paterna, Apt.~22085, E-46071 Val\`encia, Spain}
\maketitle
\begin{abstract}
We discuss the potential of the Sudbury Neutrino Observatory (SNO) to 
constraint the four--neutrino mixing schemes favoured by the results 
of all neutrino oscillations experiments. 
These schemes allow simultaneous 
transitions of solar $\nu_e's$ into active $\nu_\mu$'s, $\nu_\tau$'s 
and sterile $\nu_s$ controlled by the additional parameter 
$\cos^2(\vartheta_{23}) \cos^2(\vartheta_{24})$ and they contain as 
limiting cases the pure $\nu_e$--active and $\nu_e$--sterile neutrino 
oscillations. We first obtain the solutions allowed by the existing data 
in the framework of the  BP00 standard solar model and quantify the
corresponding predictions for the Charged Current  and
the Neutral Current to Charged Current event ratios at SNO in the
different allowed regions as a function  of the active--sterile admixture.
Our results show that some information on the 
value of $\cos^2(\vartheta_{23}) \cos^2(\vartheta_{24})$ 
can be obtained by the first SNO measurement of the CC  
ratio, while considerable improvement on the knowledge of this mixing 
will be achievable after the measurement of the NC/CC ratio. 
\end{abstract}
\end{titlepage}
\section{Introduction} 
The Sudbury Neutrino Observatory \cite{sno} is a second generation
water Cerenkov detector using 1000 tonnes of heavy water, D$_2$O, 
as detection medium. SNO was designed to address the problem of the
deficit of solar neutrinos observed previously in the 
Homestake~\cite{chlorine}, Sage~\cite{sage}, Gallex+GNO~\cite{gallex,gno} ,
Kamiokande~\cite{kamioka} and Super--Kamiokande~\cite{superk,suzuki} 
experiments, 
by having sensitivity to all flavours of neutrinos and not just to
$\nu_e$, allowing for a model independent test of the oscillation 
explanation of the observed deficit. 

Such sensitivity can be achievable because energetic neutrinos can interact
in the D$_2$O of SNO via three different reactions. 
Electron neutrinos may interact via the Charged Current (CC) reaction
\begin{equation}
\nu_e + d \to p + p + e^-\ ,
\label{CCreac}
\end{equation}
with an energy threshold of several MeV. 
All non-sterile neutrinos may also interact via Neutral Current (NC)
\begin{equation}
\label{eq:nunc}
\nu_x + d \to n + p + \nu'_x  \ ,\quad (x=e,\mu,\,\tau),
\label{NCreac}
\end{equation}
with an energy threshold of 2.225 MeV. With smaller cross section the
non-sterile neutrinos can also interact via Elastic Scattering (ES) 
$\nu_x + e^- \to \nu'_x + e^- $.

The main objective of SNO is to measure the ratio of NC/CC events. 
In its first year of operation SNO is concentrating on the measurement
of the CC reaction rate while in a following phase, after the addition
of MgCl$_2$ salt to enhance the NC signal, it will also perform a precise
measurement of the NC rate. 
It is clear that a cross-section-normalized and acceptance-corrected ratio
higher than 1 would strongly indicate the oscillation of $\nu_e$ 
into $\nu_\mu$ and/or $\nu_\tau$. On the other hand a deficit on both
CC and NC leading to a normalized NC/CC ratio $1$, can
only be made compatible with the oscillation hypothesis if  
$\nu_e$ oscillates in to a sterile neutrino. 

There are several detailed studies in the literature of the potential of
the SNO experiment to discriminate between the different oscillation
solutions to the solar neutrino problem (SNP) 
\cite{bali,snoothers,bkssno,ournew}. 
Most of these studies have been performed in the framework of 
oscillations between two neutrino states where $\nu_e$ oscillates
into either an active, $\nu_e\to\nu_a$,  or a sterile, $\nu_e\to\nu_s$, 
neutrino channel. On the other hand, 
once the possibility of a sterile neutrino is considered, 
these two scenarios are only limiting cases of the most general 
mixing structure \cite{DGKK-99,ourfour} which permits 
simultaneous $\nu_e\to\nu_s$ and $\nu_e\to\nu_a$ oscillations. 

In this paper we study the potential of the Sudbury Neutrino 
Observatory to discriminate between active and sterile solar neutrino 
oscillations when analyzed in the framework of four--neutrino mixing.
We consider those four--neutrino schemes favoured by considering together
with the solar neutrino data, the results of the two additional evidences 
pointing out towards the existence of neutrino masses and mixing: 
the atmospheric neutrino data \cite{atmos} and the LSND results \cite{lsnd}. 
We concentrate on two SNO measurements: the first expected result on the 
CC ratio and the expected to be most sensitive, the ratio of NC/CC.  
The measurement of other observables,
such as the recoil energy spectrum of the CC events and the zenith
angular dependence \cite{bali,snoothers,bkssno,ournew}
can provide important information to distinguish
between the different allowed regions for $\nu_e$--active oscillations 
but they are not expected to be very sensitive as discriminatory between
the active and sterile oscillations.    
 
The outline of the paper is the following. For the sake of completeness we 
begin by discussing in  Sec.~\ref{two} the expected results when 
obtained in the pure two--neutrino oscillation hypothesis.
In Section~\ref{fourana} we determine the presently allowed regions for the
oscillation solutions to the SNP  in the framework 
of four--neutrino mixing. In Sec.~\ref{foursno} we present the results
of the expected CC and NC/CC rates for the different solutions and
quantify the attainable sensitivity to the additional mixing controlling
the admixture of active--sterile entering into the solar 
neutrino oscillations. Finally in Sec.~\ref{discuss} we summarize 
our conclusions.

\section{Two--Neutrino Mixing: Allowed Regions and Predictions for SNO}
\label{two}
We first describe the results of the analysis of the solar neutrino 
data in terms of $\nu_e$ oscillations into either active or sterile neutrinos.
We determine the allowed range of oscillation parameters using
the total event rates of the Chlorine \cite{chlorine}, Gallium 
\cite{sage,gallex,gno} and Super--Kamiokande \cite{superk,suzuki} 
(corresponding to  the 1117 days data sample) experiments. For the
Gallium experiments we have used the weighted average of the results
from GALLEX+GNO and SAGE detectors. We have also include the 
Super--Kamiokande electron recoil energy spectrum measured separately
during the day and night periods. 
This will be referred in the following as the day--night spectra data which
contains $18 + 18$ data bins. 
The analysis includes the latest standard solar model fluxes, 
BP00 model~\cite{bp00}, with updated distributions for neutrino 
production points and solar matter density. 
For details on the statistical analysis applied to the different
observables we refer to Ref.~\cite{ourtwo,nu2000}. 
Nevertheless, 
two comments on the statistical analysis are in order:\\
--- In the present analysis we also include the contribution to the 
theoretical errors of the event rates arising from the small uncertainty 
in the measured $S_0$--factor for the reaction  $^{16}$O(p,$\gamma$)$^{17}$F 
which is new in the BP00 model as discussed in Ref.~\cite{bp00}.
Following the standard procedure \cite{lisirates}, we include this new source 
of uncertainty for the rates, that we denote as $C_F$,  by adding a new 
fractional $1\sigma$ uncertainty $\Delta ln X_{C_F} = 0.18$. 
Since this uncertainty affects in direct proportion to the  $^{17}$F flux we
correspondingly add a  new line $\alpha_{iC_F}=
\partial \Phi_i/\partial X_{C_F}$  
to the response matrix, with values $\alpha_{\,^{17}F, C_F}=1$ 
and $\alpha_{i C_F}=0$ for all other fluxes. \\
-- In the analysis of the day--night spectrum data we include the correlation 
between the systematic errors of the day and night bins which were 
conservatively ignored in Ref.\cite{nu2000}. 
Thus, we use the correlation matrix:
\begin{equation}
\sigma^2_{ij}=\delta_{ij}(\sigma^2_{i,stat}+\sigma^2_{i,uncorr})+
\sigma_{i,exp} \sigma_{j,exp}+\sigma_{i,cal}\sigma_{j,cal}
\end{equation}
where i and j run from 1 to 36 bins in the day--night spectra data.
$\sigma_{i,stat}$ is the statistical error, 
and $\sigma_{i,uncorr}$ is the error due to uncorrelated systematic
uncertainties. $\sigma_{i,exp}$ and $\sigma_{i,cal}$ are the 
correlated errors due to correlated systematic experimental uncertainties 
and the calculation of the expected spectrum respectively 
(see Ref.~\cite{ourtwo} for details).  
The addition of the correlations between the errors for the day and night
bins, which more properly takes into account 
the day--night information, leads to stronger constraints 
on the regeneration region. 

With all this we obtain that using the predicted fluxes from the BP00 model 
the $\chi^2$  for the total event rates is $\chi^2_{SSM}=56$ for 3 d.~o.~f.
This means that the SSM together with the SM of particle interactions
can explain the observed data with a probability lower than
$5\times 10^{-12}$. 

\begin{figure}
\begin{center} 
\mbox{\epsfig{file=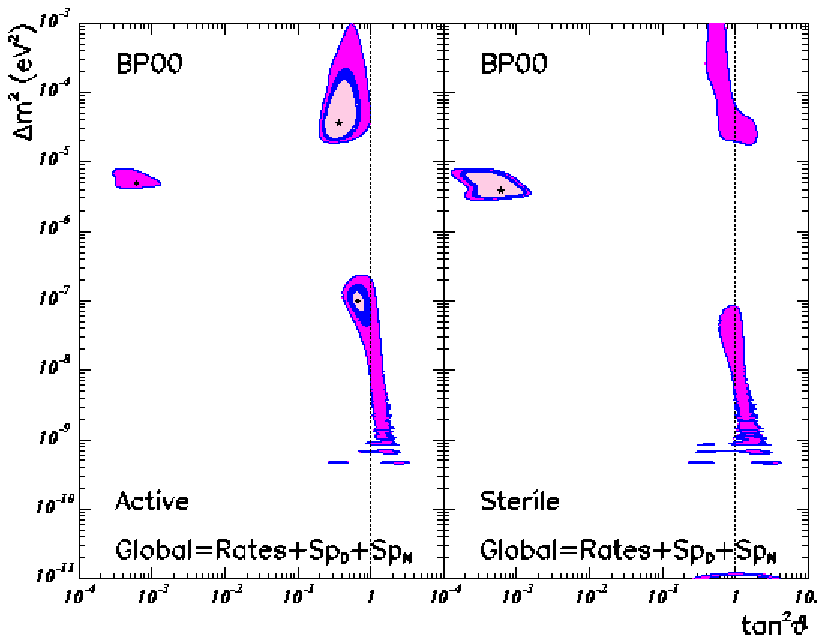,height=0.4\textheight}} 
\end{center} 
\caption{90, 95 and 99 \%CL allowed regions in the two--neutrino 
oscillation  scenario from the global analysis of solar neutrino data 
including the total 
measured rates and the Super--Kamiokande measured spectrum at day and night. 
The global 
minimum is marked with a star while the local minima are denoted with a dot.}
\label{2frspdn}
\end{figure}

The allowed regions in the oscillation parameter space are shown in 
Fig.~\ref{2frspdn}. We present them in 
the full parameter space for oscillations including both
MSW\cite{msw} and vacuum\cite{vacuum}  oscillations, 
as well as quasi-vacuum\cite{qvo} oscillations 
(QVO) and matter effects 
for mixing angles in the second octant 
(the so called dark side \cite{dark,ourfour,nu2000}). 
In the case of $\nu_e$--active neutrino oscillations we 
find that the best--fit point is obtained for the 
LMA solution. There are two more local minima of $\chi^2$ in the MSW region: 
the SMA and LOW solutions. Notice also
that LOW and QVO regions are connected at the 99 \%CL and they 
extend into the second octant so maximal mixing is allowed at 99 \% CL 
for $\Delta m^2$ in what we define as the LOW--QVO region.  

Following the standard procedure,  
the allowed regions are defined in terms of shifts of the
$\chi^2$ function {\sl with respect to the global
minimum in the plane}. Defined this way, the size of a region depends on the
{\sl relative} quality of its local minimum with respect to the global minimum
but from the size of the region we cannot infer the actual {\sl absolute} 
quality of the description in each region. In order to give this information 
we list in Table ~\ref{minima} the goodness of the fit (GOF) 
for each solution obtained from the value of $\chi^2$ at the different minima. 

For oscillations into sterile neutrinos the global minimum lies in 
the SMA solution. As seen in Fig.~\ref{2frspdn} we find that with the 
present data and using the criteria explained above, there are also 
allowed solutions for sterile neutrinos in the LMA and LOW--QVO regions 
at 99 \%CL once the day--night spectra data is included. 
We consider, however, that they are not acceptable solutions as 
their fit to the global rates is really poor with a probability of 
acceptance less than 0.004 
\footnote{Marginally allowed VO solutions 
were also possible (see for instance Ref.~\cite{goswani}) 
with last year data sample but they are now ruled out.}. 
The differences between both oscillation scenarios (active and sterile) 
can be easily understood. Unlike active neutrinos which lead to events in the
Super--Kamiokande  detector by interacting via NC with the
electrons, sterile neutrinos do not contribute to the Super--Kamiokande
event rates.  Therefore a larger survival probability for $^8$B
neutrinos is needed to accommodate the measured rate. As a consequence
a larger contribution from $^8$B neutrinos to the Chlorine and Gallium
experiments is expected, so that the small measured rate in Chlorine
can only be accommodated if no $^7$Be neutrinos are present in the
flux. This is only possible in the SMA solution region, since in the
LMA and LOW regions the suppression of $^7$Be neutrinos is not enough.
Notice also that the SMA region for oscillations into sterile neutrinos is 
slightly shifted downwards as compared with the active case. This is due
to the small modification on the neutrino survival probability induced
by the different matter potentials. The matter potential for sterile
neutrinos is smaller than for active neutrinos due to the negative NC
contribution proportional to the neutron abundance. 
For this reason the resonant condition for sterile neutrinos is achieved at
lower $\Delta m^2$. On the other hand, the flatter spectrum, is better 
fitted in both LMA and LOW regions independently of the active or sterile 
nature of the neutrino. This leads to the improvement of the quality of the
description for these solutions for both active and sterile neutrinos.
However, as mentioned above, for the analysis of the total rates 
these LMA and LOW solutions give a very bad fit in the sterile case and 
we decide not to consider them in the following. Also, as we will 
see in next section, when the analysis is performed in the framework 
of four--neutrino oscillations those large mixing solutions for 
sterile neutrinos do not appear.

Next we quantify the predictions for the SNO observables in the allowed
regions discussed above. 
The total number of events in the CC reaction at SNO can be obtained as
\begin{eqnarray} 
N^{th}_{CC} & = & \sum_{k=1,2} \phi_k 
\int\! dE_\nu\, \lambda_k (E_\nu) \sigma_{CC}(E_\nu)  \langle  
P_{\nu_e\to\nu_e}\rangle \label{eventscc}
\end{eqnarray}   
where $E_\nu$ is the neutrino energy, $\phi_k$ are the total neutrino
 $^8$B and hep fluxes, $\lambda_k$ is the neutrino energy spectrum 
(normalized to 1) and $\langle P_{\nu_e\to\nu_e} \rangle$ is the 
time--averaged $\nu_e$ survival probability for oscillations into
either active or sterile neutrinos. Here $\sigma_{CC}$ is 
the $\nu$d CC cross section computed from the corresponding 
differential cross sections folded with the finite energy resolution 
function of the detector and integrated over the electron recoil energy: 
\begin{equation} 
\sigma_{CC}(E_\nu)=\int_{T_{\text {th}}}\!dT 
\int \!dT'\,Res(T,\,T')\,\frac{d\sigma_{CC}(E_\nu,\,T')}{dT'}\ , 
\label{sigma} 
\end{equation} 
 where T and T' are the {\it measured} and the {\it true} kinetic 
energy of the recoil electrons 
and $T_{\text {th}}$ indicates the threshold expected from the experiment.
The resolution function $Res(T,\,T')$ is of the form~\cite{bali}: 
\begin{equation} 
Res(T,\,T') = \frac{1}{\sqrt{2\pi}(0.348 \sqrt{T'/\text{MeV}})}\exp 
\left[-\frac{(T-T')^2}{0.242\,T'\,{\text {MeV}}}\right]\ , 
\end{equation} 
and we take the differential cross section $d\sigma_{CC}(E_\nu,\,T')/dT'$  
from~\cite{kubodera}. For definiteness,we adopted the most optimistic 
total energy threshold $E_{th}=5~MeV$ ($T_{th}=E_{th}-m_e$).

Correspondingly, the total number of events in the NC reaction at SNO 
is obtained as
\begin{eqnarray} 
N^{th}_{NC} & = & \sum_{k=1,2} \phi_k 
\int\! dE_\nu\, \lambda_k (E_\nu) \sigma_{NC}(E_\nu)  (\langle  
P_{\nu_e\to\nu_e}\rangle +\langle  
P_{\nu_e\to\nu_a}\rangle)
\label{eventsnc}\end{eqnarray}   
where $\sigma_{NC}$ is the $\nu$d NC cross section from~\cite{kubodera} 
and $\langle P_{\nu_e\to\nu_a} \rangle$ is the time--averaged probability
of oscillation into any other active neutrino. In the case that 
$\nu_e$ oscillates only into active neutrinos 
$\langle P_{\nu_e\to\nu_e}\rangle +\langle P_{\nu_e\to\nu_a}\rangle$=1 
and $N^{th}_{NC}$ is a constant.

In order to cancel out all energy independent efficiencies and normalizations 
we will use the ratio:
\begin{eqnarray} 
R^{th}_{CC} & = & \frac{N^{th}_{CC}}{N^{SSM}_{CC}}\equiv{\mbox{[CC]}} 
\label{ratescc}
\end{eqnarray}   
where $N^{SSM}_{CC}$ is the predicted number of events in the case of 
no oscillations. The equivalent expression for the NC ratio
\begin{eqnarray} 
R^{th}_{NC} & = & \frac{N^{th}_{NC}}{N^{SSM}_{NC}}\equiv{\mbox{[NC]}}  
\label{ratesnc}
\end{eqnarray}   
Out of those ratios one can compute the double ratio 
$\frac{R^{th}_{NC}}{R^{th}_{CC}}\equiv$ [NC]/[CC] for which the largest 
sources of uncertainties cancel out ~\cite{bkssno}. As it was shown 
in Ref.~\cite{kubodera}, the ratio between the NC and 
CC reaction cross sections is extremely stable against 
any variations of the inputs of the calculations. The expected 
total uncertainties for the [CC] ratio and the  [NC]/[CC] ratio are 6.7 \% and 
3.6 \% respectively assuming 5000 CC events and 1219 NC events~\cite{bkssno}.

\begin{figure}
\begin{center} 
\mbox{\epsfig{file=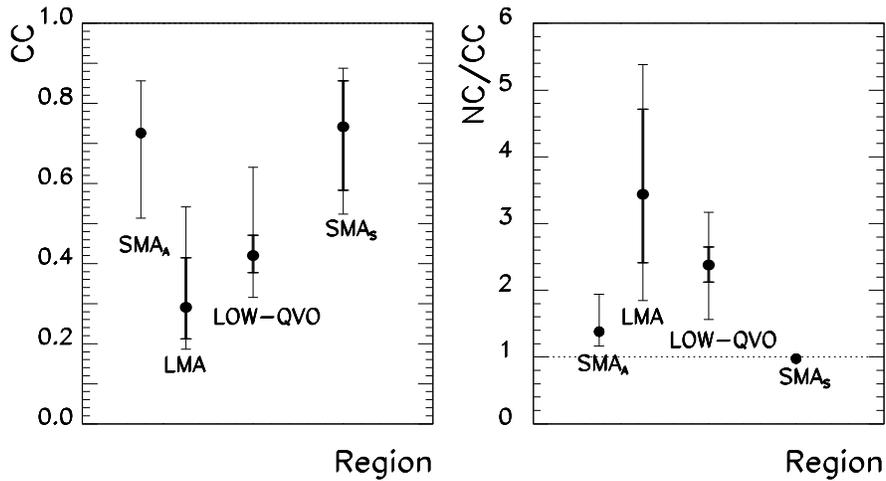,height=0.3\textheight}} 
\end{center} 
\caption{[CC] and [NC]/[CC] predictions at SNO for the allowed regions 
in the two--neutrino mixing  scenarios obtained from the global analysis 
of solar neutrino data at 90 \% and 99 \%CL.}  
\label{2fsno}
\end{figure}

In Fig.~\ref{2fsno} we show the predicted [CC] and  [NC]/[CC] ratios 
for the allowed regions in the two flavour analysis. The dots 
correspond to the local best fit points and the error bars show the
range  of predictions for the points inside the 90 and 99 \%CL allowed regions.
The mapping of the regions onto these bars can be easily understood from 
the behaviour of the probability for the different solutions: \\
(a) For oscillations into active neutrinos the [NC]/[CC] ratio is 
simply the inverse of the [CC] prediction. Therefore 
\begin{itemize}
\item  In the SMA region smaller mixing angles are mapped onto  
higher (lower) values of [CC]  ([NC]/[CC]) ratio. One may notice that 
the prediction for the 
[CC] rate for the global best fit point (0.72) is larger than the measured 
rate at Super--Kamiokande. This is due to the nearly flat spectrum 
at Super--Kamiokande which implies that the best fit point in the global 
analysis corresponds to a smaller mixing angle than the best fit point 
for the analysis of rates only.
\item  In the LMA region, the lower $\Delta m^2$ and $\theta$ values 
are mapped onto higher (lower) [NC]/[CC]  ([CC]) ratios and viceversa.
\item  In the LOW region the higher (lower) [NC]/[CC]  ([CC]) ratio occurs 
for smaller $\theta$ and higher $\Delta m^2$. 
\end{itemize}
(b) For the sterile case, the best fit point in SMA occurs at 
lower $\Delta m^2$ than in the active case and this produces a higher 
prediction for the [CC] ratio (0.74). The [NC]/[CC] ratio 
takes an almost constant value very close to one (0.98 in the best fit point),
since both numerator and denominator are proportional to 
$\langle P_{\nu_e\to\nu_e}\rangle$. 
It is smaller than one because for the SMA solution the probability increases 
with energy in the range of detection at SNO and the threshold for the 
NC reaction is below the one for the CC one.

For the sake of consistency we have checked that our results agree perfectly 
with those in Ref.~\cite{bkssno} when comparing the same points in the 
parameter space. However a careful reader may notice that the predictions 
at the best fit points and ranges in each region displayed in Fig.~\ref{2fsno} 
are slightly different of those in Ref.~\cite{bkssno}. The 
difference is due to two factors. First, 
the allowed regions are defined
in a different way. In Ref.~\cite{bkssno} departures from the standard solar 
model in the boron flux normalization are allowed and moreover the 
regions are defined in terms of shifts of the $\chi^2$ function 
{\sl with respect to the local minimum in the corresponding region}. 
Second, the inclusion of the updated data, mainly the 
Super--Kamiokande day--night spectra, lowers the value of $\tan^2 \theta$ 
for the best bit point in the SMA region by a factor ~2 and 
increases $\Delta m^2$ for the best fit point in LMA by a factor ~1.5. 

What we see from these results is that while the data on [CC]  can give 
a hint towards large or small mixing solutions, it  will be hard to 
distinguish active from sterile oscillations on the only bases of this
measurement. This is not the case for the [NC]/[CC] ratio where 
both scenarios appear nicely separated. It is not hard to foresee 
from these results that from the [NC]/[CC] measurement SNO will be able to 
constraint the additional mixings in the four--neutrino scenario which 
describe the admixture of active and sterile oscillations. This is the 
main point in this paper.

\section{Allowed Four--neutrino Mixing Parameters}
\label{fourana}
Together with the results from the solar neutrino  
experiments we have two more evidences pointing out towards the existence of  
neutrino masses and mixing: the atmospheric neutrino data \cite{atmos} 
and the LSND results \cite{lsnd}. All these experimental results can be 
accommodated in a single neutrino oscillation framework only if there 
are at least three different scales of neutrino mass-squared differences. 
The simplest case of three independent mass-squared differences 
requires the existence of a light sterile neutrino, 
{\it i.e.} one whose interaction with 
standard model particles is much weaker 
than the SM weak interaction, 
so it does not affect the invisible Z decay  
width, precisely measured at LEP.

There are six possible four--neutrino schemes that can accomodate all these
evidences. They can be divided in two classes: 3+1 and 2+2. In the 
3+1 schemes there is a group of three neutrino masses separated from an 
isolated mass by a gap of the order of 1eV which gives the mass-squared 
difference responsible for the short-baseline oscillations observed in the 
LSND experiment. In 2+2 schemes there are two pairs of close masses separated 
by the LSND gap. We have ordered the masses in such a way that
in all these schemes $ \Delta{m}^2_{\text{sun}} = \Delta{m}^2_{21} $
produces solar neutrino oscillations
and $ \Delta{m}^2_{\text{LSND}} = \Delta{m}^2_{41} $
(we use the common notation $\Delta{m}^2_{kj} \equiv m_k^2 - m_j^2$).  
3+1 schemes are disfavoured by experimental data with respect to the
2+2 schemes \cite{BGG-AB,Barger-variations-98} but they  
are still marginally allowed \cite{3+1}. 

In any of these four-neutrino schemes the flavour neutrino fields 
$\nu_{\alpha L}$ (we choose $\alpha=e,s,\mu,\tau$) 
are related to the fields $\nu_{kL}$ of neutrinos with masses $m_k$ 
by a rotation $U$. $U$ is a $4{\times}4$ unitary mixing matrix, 
which contains, in general, 
6 mixing angles and 3 CP violating phases (3 additional phases appear
for Majorana neutrinos but they are irrelevant for oscillations).
We neglect here the CP phases, which, in the schemes considered, 
are irrelevant for solar neutrinos because their effect is washed out 
by averaging over neutrino energy and distance.
Existing bounds from negative searches for neutrino oscillations performed
at colliders as well as reactor experiments, in particular 
the negative results of the Bugey \cite{bugey}  
and CHOOZ \cite{chooz} $\bar\nu_e$ disappearance experiment, impose severe 
constrains on the possible mixing structures 
for the four--neutrino scenario. In particular they imply that the matrix 
elements $U_{e3}$ and $U_{e4}$ are very small 
\cite{BGG-AB,Barger-variations-98,GL}. 
As a consequence, for any of these four--neutrino schemes, either 2+2 or
3+1, only four mixing angles are relevant in the study of solar neutrino 
oscillations~\cite{DGKK-99,ourfour,GL} and the $U$ matrix can be written as
\begin{equation}
U=
\left( \begin{array}{cccc} 
c_{12}
& s_{12}
& 0
& 
0
\\ 
- s_{12} c_{23} c_{24}
& 
c_{12} c_{23} c_{24}
& 
s_{23} c_{24}
& 
s_{24}
\\
s_{12}
( c_{23} s_{24} s_{34}
+ s_{23} c_{34} )
& 
- c_{12}
( s_{23} c_{34}
+ c_{23} s_{24} s_{34} )
& 
c_{23} c_{34}
- s_{23} s_{24} s_{34}
& 
c_{24} s_{34}
\\
s_{12}
( c_{23} s_{24} c_{34}
- s_{23} s_{34} )
& 
c_{12}
( s_{23} s_{34}
- c_{23} s_{24} c_{34} )
& 
-
( c_{23} s_{34}
+ s_{23} s_{24} c_{34} )
& 
c_{24} c_{34}
\end{array}
\right)
\,,
\label{U-sun}
\end{equation}
where
$\vartheta_{12}$,
$\vartheta_{23}$,
$\vartheta_{24}$,
$\vartheta_{34}$
are four mixing angles
and
$ c_{ij} \equiv \cos\vartheta_{ij} $
and
$ s_{ij} \equiv \sin\vartheta_{ij} $.

Since solar neutrino oscillations 
are generated by the mass-square difference 
between $\nu_2$ and $\nu_1$, 
it is clear from Eq.~(\ref{U-sun}) 
that the survival of solar $\nu_e$'s 
mainly depends on the mixing angle 
$\vartheta_{12}$, 
whereas the mixing angles 
$\vartheta_{23}$ and $\vartheta_{24}$ 
determine the relative amount of transitions into sterile $\nu_s$ 
or active $\nu_a$, this last one being a combination of 
$\nu_\mu$ and $\nu_\tau$ controlled by the mixing angle $\theta_{34}$. 
$\nu_\mu$ and $\nu_\tau$ 
cannot be distinguished in solar neutrino experiments, 
because their matter potential 
and their interaction in the detectors are equal, 
due only to NC weak interactions. 
As a consequence the active/sterile ratio  and the survival 
probability for solar 
neutrino oscillations do not depend on the mixing angle 
$\vartheta_{34}$, and depend on the mixing angles 
$\vartheta_{23}$ 
$\vartheta_{24}$ only through the combination 
$\cos{\vartheta_{23}} \cos{\vartheta_{24}}$. For
further details see Ref.~\cite{DGKK-99,ourfour}.
We distinguish the following limiting cases: \\
$\bullet$ $\cos{\vartheta_{23}} \cos{\vartheta_{24}} = 0$ corresponding to the limit of pure two-generation 
$\nu_e\to\nu_a$ transitions.\\
$\bullet$ $\cos{\vartheta_{23}} \cos{\vartheta_{24}} = 1$ 
for which we have the limit of 
pure two-generation 
$\nu_e\to\nu_s$ transitions.\\
$\bullet$ If $\cos{\vartheta_{23}} \cos{\vartheta_{24}} \neq 1$, 
solar $\nu_e$'s can transform 
in the linear combination $\nu_a$ of active $\nu_\mu$ and $\nu_\tau$. 

In the general case of simultaneous $\nu_e\to\nu_s$ and $\nu_e\to\nu_a$ 
oscillations 
the corresponding probabilities are given by~\cite{DGKK-99,ourfour}
\begin{eqnarray} 
&& 
P_{\nu_e\to\nu_s} 
= 
c^2_{23} c^2_{24} 
\left( 1 - P_{\nu_e\to\nu_e} \right) 
\,, 
\label{Pes} 
\label{Pea} 
\\ 
&& 
P_{\nu_e\to\nu_a} 
= 
\left( 1 - c^2_{23} c^2_{24} \right) 
\left( 1 - P_{\nu_e\to\nu_e} \right) 
\,. 
\end{eqnarray} 
where  $P_{\nu_e\to\nu_e}$ takes the standard two--neutrino oscillation
form for $\Delta m^2_{12}$ and $\theta_{12}$ but
computed with the modified matter potential  
\begin{equation} 
A 
\equiv 
A_{CC} + c^2_{23}c^2_{24} A_{NC} 
\,. 
\label{A} 
\end{equation} 
Thus the analysis of the solar neutrino data in the
four--neutrino mixing schemes is equivalent to the two--neutrino
analysis but taking into account that the parameter space is now 
three--dimensional $(\Delta m^2_{12},\tan^2\vartheta_{12}, 
\cos^2{\vartheta_{23}} \, \cos^2{\vartheta_{24}})$. 
We want to stress that, although originally this derivation 
was performed in the framework of the 2+2 schemes~\cite{DGKK-99,ourfour},
it is equally valid for the 3+1 ones \cite{GL}.

\begin{figure}
\begin{center} 
\mbox{\epsfig{file=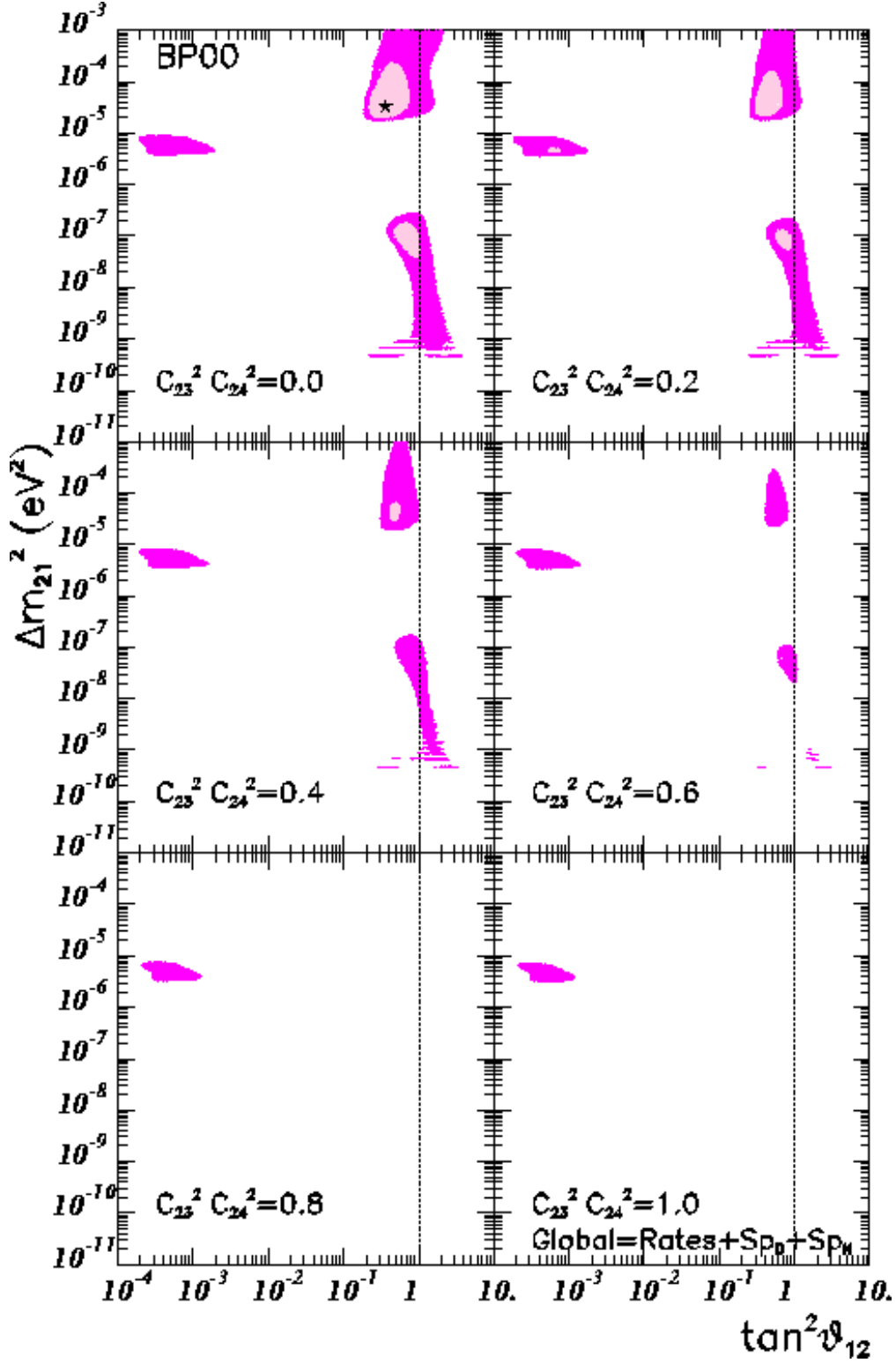,height=0.9\textheight}} 
\end{center} 
\caption{Results of the global analysis for the allowed regions in   
$\Delta{m}^2_{21}$ and $\sin^2 \vartheta_{12}$  
for the four--neutrino oscillations. 
The different panels 
represent the allowed regions at 90 \% (lighter) and 99\%CL (darker).  
The best--fit point in the three parameter space is  
plotted as a star.} 
\label{four}
\end{figure}

We first present the results of the allowed regions in the three--parameter 
space for the global combination of observables. 
Notice that since the parameter space is 3--dimensional the allowed regions  
for a given CL are defined as the set of points satisfying  
the condition
$\chi^2(\Delta m_{12}^2,\vartheta_{12},c_{23}^2c_{24}^2)
-\chi^2_{min}\leq \Delta\chi^2 \mbox{(CL, 3~dof)} $
where, for instance, $\Delta\chi^2($CL, 3~dof)=6.25, 7.83, and 11.36 for 
CL=90, 95, and 99 \% respectively. 
In Figs.~\ref{four} we plot the  
sections of such volume in the plane 
($\Delta{m}^2_{21},\tan^2(\vartheta_{12})$) for different values of 
$c_{23}^2c_{24}^2$.  The global minimum used in the construction of the 
regions lies in the LMA region and for pure $\nu_e$--active oscillations, 
$c_{23}^2c_{24}^2=0$. 

As seen in  Fig.~\ref{four} the SMA region is always a valid solution  
for any value of $c_{23}^2c_{24}^2$ at 99\% CL (the same is true at
95\% CL). This is expected as  
in the two--neutrino oscillation picture this solution holds both  
for pure $\nu_e$--active and pure $\nu_e$--sterile oscillations.
Notice, however, that the statistical analysis is different: 
in the two--neutrino picture the pure $\nu_e$--active and $\nu_e$--sterile 
cases are analyzed separately, 
whereas in the four--neutrino picture they are taken into account 
simultaneously in a consistent scheme. Since the 
GOF of the SMA solution for pure $\nu_e$--sterile oscillations is worse than
for SMA pure active oscillations (as discussed in the previous section), 
the corresponding allowed region is smaller because they are now defined 
with respect to a common minimum. Also, we notice, that for the SMA
solution the best scenario is a non-zero admixture between active and
sterile oscillations. For this reason this solution is allowed
at a CL better than 90 \% only in the range 
$0.11 \leq c_{23}^2c_{24}^2 \leq 0.31$. 

On the other hand, the LMA and LOW--QVO solutions disappear for 
increasing values of the mixing $c_{23}^2c_{24}^2$. 
We list in Table \ref{limits4} the ranges of
$c_{23}^2c_{24}^2$ 
for which each of the solutions is allowed at 
a given CL. We see that at 95 \%CL the LMA solution is allowed 
for maximal active--sterile mixing $c_{23}^2c_{24}^2=0.5$ while 
at 99\%CL all solutions are possible for maximal admixture.

\section{Expected Rates at SNO in Four--Neutrino Schemes}
\label{foursno}
In this section, we present the predictions for the CC ratio and for 
the NC/CC ratio in the four--neutrino scenario previously described. 
This scenario contains as limiting cases the pure $\nu_e$--active 
and $\nu_e$--sterile neutrino oscillations. However, when comparing the 
results for both limiting cases with the ones presented 
in  Sec.~\ref{two} the reader must notice that there are  
some changes in the predicted ranges 
because the allowed regions are obtained with a different statistical 
criteria. Now, as discussed above, all the allowed regions are defined 
with respect to the same global minimum (laying in the LMA 
with $c_{23}^2c_{24}^2$=0) with 3~dof. Because of that, the predicted 
ranges in the four--neutrino scheme are wider for the pure $\nu_e$--active 
oscillations and narrower for the $\nu_e$--sterile case.

In Figs.~\ref{4fsno_sma}$-$\ref{4fsno_low} we show the results for the 
predicted [CC] ratio and [NC]/[CC] ratio for the different allowed 
regions (SMA, LMA, LOW-QVO) at 90 and 99 \%CL as a function 
of $c_{23}^2c_{24}^2$. 
The general behaviour of the dependence of the predicted ratios with 
$c_{23}^2c_{24}^2$ can be easily understood using the following 
simplified expressions obtained from Eqs.~(\ref{ratescc}), 
(\ref{ratesnc}) and (\ref{Pea}):
\begin{eqnarray}
\mbox{\rm [CC]} &\sim& 
 P_{\nu_e\to\nu_e} \; , \label{CC}\\
\frac{\mbox{\rm [NC]}}
{\mbox{\rm [CC]}} &\sim &
\frac{1 - c^2_{23} c^2_{24}(1 - P_{\nu_e\to\nu_e})}{P_{\nu_e\to\nu_e}}\;. 
\label{NCCC}
\end{eqnarray}
From Eq.~(\ref{CC}) we see that the only dependence of [CC] on 
$c_{23}^2c_{24}^2$  is 
due to the modification of the matter potential entering in  
the evolution equation (see Eq.~(\ref{A}) and discussion above)  
and it is very weak. The dependence of the allowed range
of the [CC] ratio with  $c_{23}^2c_{24}^2$ displayed in the figures
arises mainly from the variation of the size of the allowed regions. 
Alternatively following Eq.~(\ref{NCCC}) we find a stronger linear 
dependence of [NC]/[CC] on $c_{23}^2c_{24}^2$ with slope  
$(- 1 + P_{\nu_e\to\nu_e})/P_{\nu_e\to\nu_e}\sim 1-1/$[CC] 
and intercept $1/P_{\nu_e\to\nu_e}\sim 1/$[CC].
This simple description is able to reproduce the main features of our 
numerical calculations as can be seen in the figures. 

Figs.~\ref{4fsno_sma}$-$\ref{4fsno_low} contain the main quantitative
result of our analysis in the four--neutrino mixing scenario. From each 
of them it is possible to infer the allowed range of the active--sterile 
admixture, $c_{23}^2c_{24}^2$, compatible, within the expected uncertainty, 
with a given SNO measurement of the ratios. 
Also, comparing the allowed ranges for the different solutions 
one can study the potential of these measurements as discriminatory 
among the three presently allowed regions. Of course, both issues are 
not independent as we have no ``a priory'' knowledge of which is the
right solution and both must be discussed simultaneously. 
In order to do so we pass to describe and compare in detail the 
predictions in the different regions.

The results for the SMA solution are shown in Fig.~\ref{4fsno_sma}.a 
and~\ref{4fsno_sma}.b for [CC] and  [NC]/[CC] ratios respectively. 
First we notice that we find a small region allowed at 90 \%CL only for 
a non--vanishing admixture of active and sterile oscillations as mentioned 
before. In this region [CC]$\sim$0.65--0.73 and 
[NC]/[CC]$\sim$ 1.3--1.4. The predictions at 99\% range from 
[CC]$\sim$ 0.4--0.9 ([NC]/[CC]$\sim$1.1--2.5) 
for pure $\nu_e$--active scenario to 
to [CC]$\sim$ 0.59--0.85 ([NC]/[CC]$\sim$0.96--0.98) for pure 
$\nu_e$--sterile oscillations.
Thus if SNO observes a ratio [CC]$<$0.58 the 
value of $c_{23}^2c_{24}^2$ can be constrained to be smaller than 1 
disfavouring pure $\nu_e$--sterile oscillations. On the contrary 
a measurement of [CC]$\gsim$0.68 will immediately hint towards 
the SMA solution but will not provide any information on the active--sterile
admixture. Also, one must notice, that such value, although allowed
by the present global statistical analysis at 99 \%CL, will imply
a strong disagreement with the total rate event rate observed at 
Super--Kamiokande.

As seen in Fig.~\ref{4fsno_sma}.b the [NC]/[CC] ratio is more sensitive 
to the active--sterile admixture. 
To guide the eye, in the figures for the 
[NC]/[CC] ratio we plot a dotted line for the prediction in the case of 
no oscillation [NC]/[CC]=1. For any of the solutions,  
the allowed range for this ratio shows as general behaviour a decreasing 
with $c_{23}^2c_{24}^2$ due to two effects: $(i)$ the allowed regions become
smaller and $(ii)$ the prediction decreases when more sterile neutrino 
is involved in the oscillations as described in Eq.~(\ref{NCCC}). The 
measurement of higher values of this ratio will favour the four--neutrino
scenario with larger component of $\nu_e$--active oscillations. On the
other hand a measurement of  [NC]/[CC]$\sim 1$, will push the 
oscillation hypothesis towards the pure 
$\nu_e$--sterile oscillation scenario. This case will 
be harder to differentiate from the non--oscillation scenario. 
We find that with the expected sensitivity the parameter 
$c_{23}^2c_{24}^2$ is constrained to be above 0.44 at 99\% CL and that 
the pure $\nu_e$--active oscillations in the SMA region are compatible 
with [NC]/[CC]= 1 only at $\sim 5\sigma$.

The predictions for oscillation parameters in the LMA region are
shown in Figs.~\ref{4fsno_lma}.a and~\ref{4fsno_lma}.b for 
[CC] and [NC]/[CC] ratios respectively. 
The predictions at 99\% vary in the range 
[CC]$\sim$ 0.18--0.62  and [NC]/[CC]$\sim$ 1.4--5.6. 
The first thing we notice by comparing Fig.~\ref{4fsno_lma}.a
with Fig.~\ref{4fsno_sma}.a and Fig.~\ref{4fsno_low}.a 
is that the most discriminatory scenario for the [CC] rate results if 
SNO finds a small value [CC]$\sim 0.25$. This would 
significantly hint towards the LMA solution to the solar 
neutrino problem and towards and $\nu_e$--active oscillation scenario. 
First, it is well separated from the predictions for the SMA and LOW 
regions. Second, it will include as a bonus a small but measurable 
day--night asymmetry \cite{snoothers,bkssno}. And 
third it will constrain the $c_{23}^2c_{24}^2$ to a small value 
($\sim 0.2$). On the contrary the less discriminatory scenario will be a 
measurement 0.4$<$[CC]$<$0.6 where the prediction would 
be compatible with both SMA and LOW-QVO solutions and no improvement 
on our knowledge of the four--neutrino schemes is possible.
The [NC]/[CC] ratio can definitively improve the discrimination between
the different scenarios provided its measurement lies in the upper range. 
For instance a measurement of 
[NC]/[CC]$\sim$ 4 ($\pm$ 0.7 at $5\sigma$) will be 
conclusive for selecting LMA as the solution to the SNP and will imply
and upper bound on  $c_{23}^2c_{24}^2<0.3$. 

The predictions for the LOW--QVO region lie between the ones for SMA and LMA
as displayed in Fig.~\ref{4fsno_low} and therefore they are more difficult
to discriminate.  The predictions at 99\% vary in 
the range [CC]$\sim$ 0.3--0.68  and [NC]/[CC]$\sim$ 1.2--3.4. 
As a consequence we see that a low [CC] ratio but still within the 
99\% CL range allowed for this region, 0.3$<$[CC]$<$0.4, 
will constrain significantly 
the $c_{23}^2c_{24}^2$ parameter compatible with this solution but 
it  will not be distinguishable from the LMA solution unless the measured 
[CC]$<$0.3. As mentioned above, 
the [NC]/[CC] ratio will be able to differentiate the LMA and LOW-QVO 
solutions if not in the range [1.5,3]. One should also notice that for the 
upper part of this range a positive measurement of the day--night asymmetry 
and the zenith dependence~\cite{ournew} will point towards the higher 
$\Delta{m}^2$ of the LOW region as the solution.

\section{Discussion}
\label{discuss}

In this paper we have studied the potential of the Sudbury Neutrino 
Observatory to discriminate between active or sterile solar neutrino 
oscillations when analyzed in the framework of four--neutrino mixing.
We considered those four--neutrino schemes favoured by considering together
with the solar neutrino data, the results of the two additional evidences 
pointing out towards the existence of neutrino masses and mixing: 
the atmospheric neutrino data \cite{atmos} and the LSND results \cite{lsnd}. 
These schemes allow simultaneous 
transitions of solar $\nu_e's$ into active $\nu_\mu$'s, $\nu_\tau$'s 
and sterile $\nu_s$ controlled by the additional parameter 
$\cos^2(\vartheta_{23}) \cos^2(\vartheta_{24})$ and they contain as 
limiting cases the pure $\nu_e$--active and $\nu_e$--sterile neutrino 
oscillations.
The allowed solar solutions have been reanalyzed including the 
recently BP00 standard solar model and the latest solar neutrino data. 
We find that the global minimum lies in the LMA region and for pure 
$\nu_e$--active oscillations ($c_{23}^2c_{24}^2=0$). We also find that in the 
framework of four--neutrino mixing the  SMA solution 
is allowed at 90\% CL for non vanishing active--sterile mixing 
$c_{23}^2c_{24}^2$ in the range [0.11,0.31].
 
\begin{figure}
\begin{center} 
\mbox{\epsfig{file=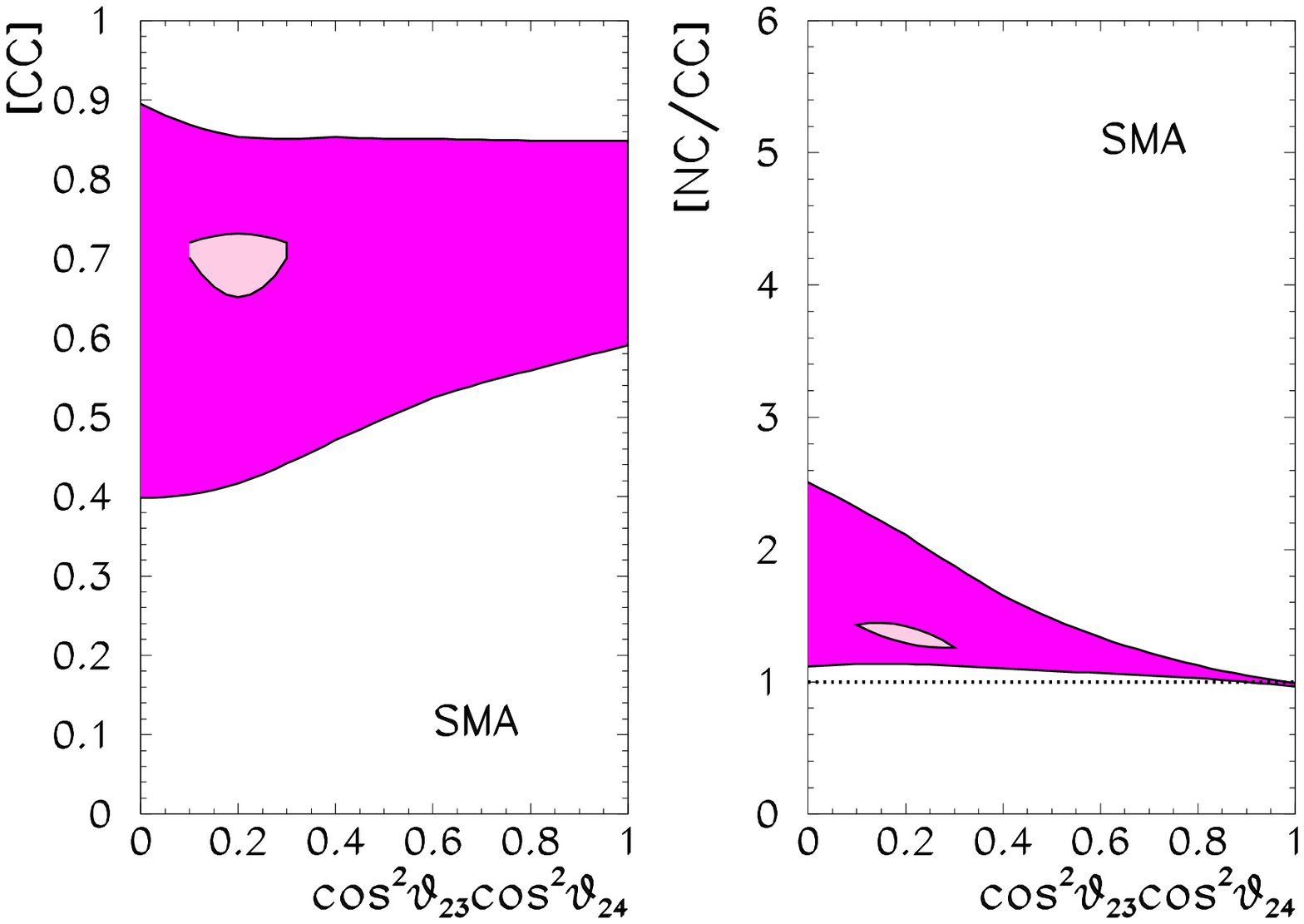,height=0.4\textheight}} 
\end{center} 
\caption{[CC] and [NC]/[CC] predictions at SNO as for the SMA region 
in the four--neutrino scenario obtained from the global analysis of 
solar neutrino data at 90 \% (lighter) and 99 \%CL (darker). The dotted 
line corresponds to the prediction in the case of no oscillations.}  
\label{4fsno_sma}
\end{figure}

\begin{figure}
\begin{center} 
\mbox{\epsfig{file=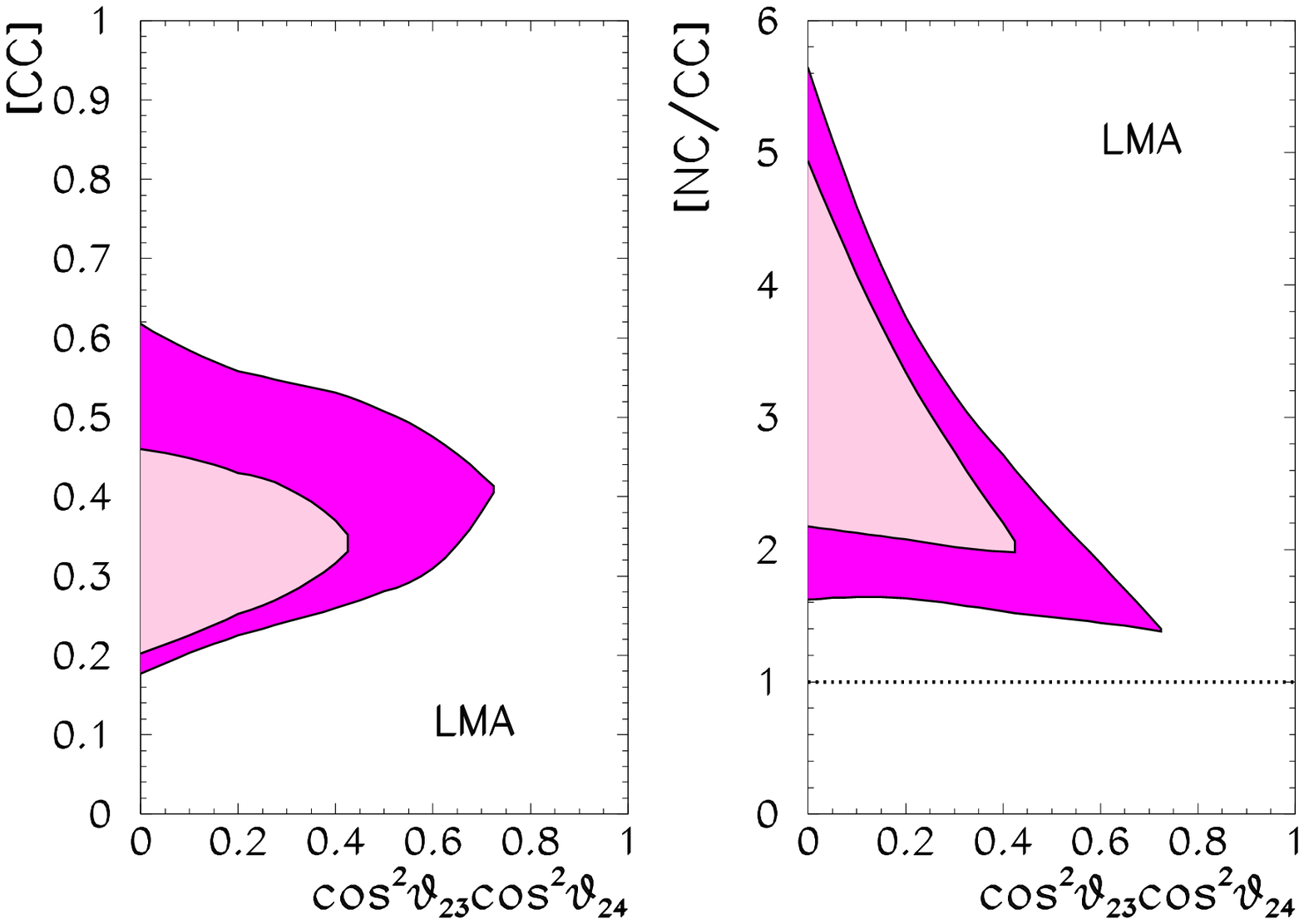,height=0.4\textheight}} 
\end{center} 
\caption{[CC] and [NC]/[CC] predictions at SNO as for the LMA region 
in the four--neutrino scenario obtained from the global analysis of 
solar neutrino data at 90 \% (lighter)  and 99 \%CL (darker). The dotted 
line corresponds to the prediction in the case of no oscillations.}  
\label{4fsno_lma}
\end{figure}

\begin{figure}
\begin{center} 
\mbox{\epsfig{file=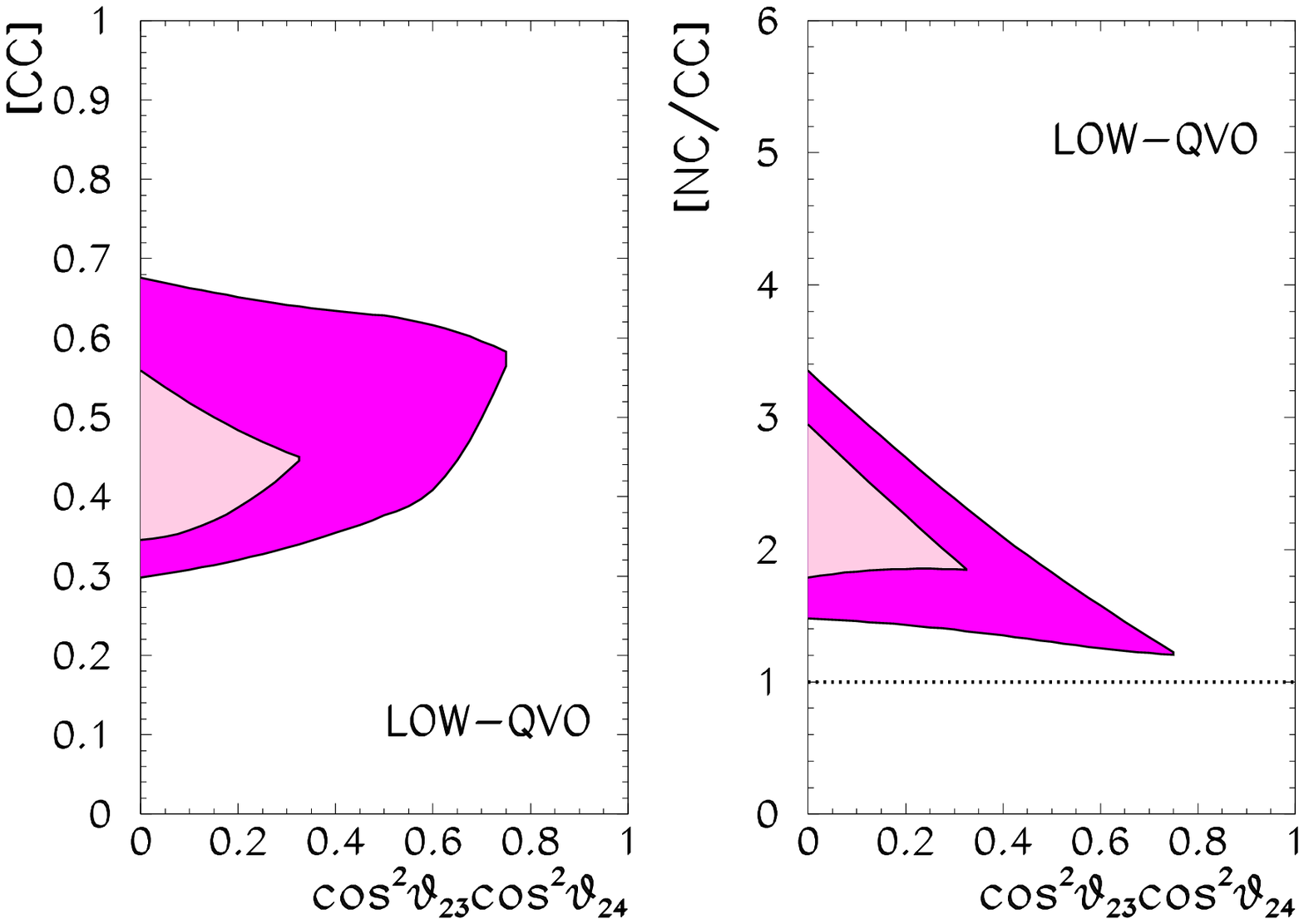,height=0.4\textheight}} 
\end{center} 
\caption{[CC] and [NC]/[CC] predictions at SNO as for the LOW region 
in the four--neutrino scenario obtained from the global analysis of 
solar neutrino data at 90 \% (lighter) and 99 \%CL (darker). The dotted 
line corresponds to the prediction in the case of no oscillations.}  
\label{4fsno_low}
\end{figure}

We concentrated on two SNO measurements: the first expected result on the 
[CC] ratio and the expected to be most sensitive to 
the active--sterile admixture, the ratio of [NC]/[CC] and evaluated
the predictions in the different regions as a function of the additional
mixing $c_{23}^2c_{24}^2$. Our results are display in 
Figs.~\ref{4fsno_sma}$-$\ref{4fsno_low}. They show that in most cases 
with the measurement of the [CC] ratio, 
it will be hard to improve the present knowledge of $c_{23}^2c_{24}^2$  
but with the precise determination of the [NC/CC] ratio at SNO, 
this parameter can be strongly constrained for some of the 
allowed solutions.
For example, we find that for the [CC] rate the most discriminatory scenario 
would be that SNO finds a small value [CC]$\sim 0.25$. This 
significantly hints towards the LMA solution to the solar 
neutrino problem and towards an active--active  oscillation scenario. 
In this case the [NC]/[CC]$\sim$ 4 ($\pm$ 0.7 at $5\sigma$) will be 
conclusive for selecting LMA as the solution to the SNP and will imply
and upper bound on  $c_{23}^2c_{24}^2<0.3$.  
Conversely, a measurement of  [NC]/[CC]$\sim 1$, 
although harder to distinguish from the non--oscillation scenario, 
will push the oscillation hypothesis towards the sterile SMA solution,  
and with the expected sensitivity a bound $c_{23}^2c_{24}^2>0.44$ 
at 99\% CL can be imposed.

\acknowledgments
We thank J.~N.~Bahcall for clarifying several aspects of BP00 standard 
solar model. We also acknowledge valuable discussions with E. Lisi on 
the correlations in the day--night spectra and with C. Giunti on the 
four--neutrino schemes. This work was supported by the spanish 
DGICYT under grants PB98-0693 and PB97-1261, by the Generalitat 
Valenciana under grant GV99-3-1-01 and by the TMR network grant 
ERBFMRXCT960090 of the European Union.


\vskip 2cm

\begin{table}
\caption{Best fit points and GOF for the allowed solutions for the
global analysis in the framework of two--neutrino mixing.}
\label{minima}
\begin{center}
\begin{tabular}{|c|c|c|c||c|}
&  \multicolumn{3}{c||}{Active}  & Sterile
\\\hline
             & SMA & LMA & LOW-QVO & SMA  \\ \hline
     $\Delta m^2$/eV$^2$ & $5.0\times 10^{-6} $  
    &  $3.7\times 10^{-5} $  
    &  $1.0\times 10^{-7} $  
    &  $3.9\times 10^{-6} $   \\ 
$\tan^2\theta$ & 0.00061     
             & 0.37   & 0.67    &0.00061 \\ \hline             
             $\chi_{min}$& 40.8  & 33.4 &  37.1  & 42.3  \\
             Prob (\%)&  27 \%  &  59 \%  & 42 \% &  22 \%     
\end{tabular}
\end{center}
\end{table} 
\begin{table}
\caption{Allowed ranges of $c^2_{23} c^2_{24}$ at 90\%, 95\%, 
and 99 \% CL for the different solutions to the solar neutrino problem.}  
\label{limits4}
\begin{center}
\begin{tabular}{|c|ccc|}
CL  & SMA & LMA & LOW-QVO \\ \hline
90  & [0.11,0.31] & [0,0.43] & [0,0.32]  \\
95  & [0,1]  & [0,0.52] & [0,0.44]  \\
99  & [0,1]  & [0,0.72] & [0,0.76] 

\end{tabular}
\end{center}
\end{table}
\end{document}